\documentclass[preprint,secnumarabic,amssymb, nobibnotes, aps, pr,superscriptaddress]{revtex4-2}
\usepackage{subfigure}
\usepackage{amsmath}
\usepackage{graphicx}
\usepackage{dcolumn}
\usepackage{bm}
\usepackage[utf8]{inputenc}
\usepackage{textcomp}
\usepackage{amsmath}
\usepackage[version=4]{mhchem}
\usepackage{siunitx}
\usepackage{caption}
\usepackage{soul}
\usepackage{booktabs}
\soulregister\cite7 
\soulregister\citep7 
\soulregister\citet7 
\soulregister\ref7 
\title{Adaptive extended Kalman filter and laser link acquisition in the detection of gravitational waves in space.}

\def\ot{\scriptscriptstyle{12}}
\def\op{\scriptscriptstyle{13}}

\begin{document}
	
	

	\title{Adaptive extended Kalman filter and laser link acquisition in the detection of gravitational waves in space.}

	\author{Jinke Yang}	\affiliation{Shanghai Institute of Technical Physics, Chinese Academy of Sciences, Shanghai, 200083, China.}
	\affiliation{%
		University of Chinese Academy of Sciences, Beijing, 100049, China.}%
	\author{Yong Xie}%
	\affiliation{%
		Shanghai Institute of Technical Physics, Chinese Academy of Sciences, Shanghai, 200083, China.}%
	\author{Yidi Fan}%
	\affiliation{%
		Key Laboratory for Satellite Digitalization Technology, Chinese Academy of Sciences, Shanghai 201203, China}%
	\author{Pengcheng Wang}%
	\affiliation{%
		Innovation Academy for Microsatellites, Chinese Academy of Sciences, Shanghai, 201203, China.}%
	\affiliation{%
		University of Chinese Academy of Sciences, Beijing, 100049, China.}%
	\author{Haojie Li}%
	\affiliation{%
		Hangzhou Institute for Advanced Study, UCAS, Hangzhou 310024, China.}%
	\author{Jianjun Jia}%
	\affiliation{%
		Shanghai Institute of Technical Physics, Chinese Academy of Sciences, Shanghai, 200083, China.}%
	\affiliation{%
		University of Chinese Academy of Sciences, Beijing, 100049, China.}%
	\author{Yucheng Tang}%
	\email{tangyucheng@mail.sitp.ac.cn}
	\affiliation{%
		Shanghai Institute of Technical Physics, Chinese Academy of Sciences, Shanghai, 200083, China.}%
	\author{Yun Kau Lau}%
	\email{lau@amss.ac.cn}
	\affiliation{Shanghai Institute of Technical Physics, Chinese Academy of Sciences, Shanghai, 200083, China.}
	\affiliation{
		Institute of Applied Mathematics, Morningside Center of Mathematics, LSSC, Academy of Mathematics and System Science, Chinese Academy of Sciences, 55, Zhongguancun Donglu, Beijing, 100190, China.	}

	\begin{abstract}
		An alternative, new laser link acquisition scheme for the triangular constellation of spacecrafts(SCs) in deep space in the detection of gravitational waves is considered. In place of a wide field CCD camera in the initial stage of laser link acquisition adopted in the conventional scheme, an extended Kalman filter based on precision orbit determination is incorporated in the point ahead angle mechanism (PAAM) to steer the laser beam in such a way to narrow the uncertainty cone and at the same time  avoids the heating  problem generated by the CCD camera. Quadrant photodetector (QPD) based on Differential Phase Shift (DPS) with higher dynamic range than that of DWS is then employed as readout for the laser beam spot. The conventional two stages (coarse acquisition and fine acquisition) are integrated into a single control loop. The payload structure of the ATP control loop is simplified and numerical simulations, based on a colored measurement noise model that closely mimics the prospective on-orbit conditions, demonstrate that the AEKF significantly reduces the initial uncertainty region by predicting the point ahead angle (PAA) even when the worst case scenario in SC position (navigation) error is considered. 
	\end{abstract}
	
	\maketitle
	
	
	\section{Introduction}
	
	In the detection of gravitational waves in space, a triangular constellation of SCs with laser links among them serve as laser interferometer to detect the picometer scale miniature change  in armlength between two SCs generated by variation in the spacetime curvature of a gravitational wave sources\cite{Taijibrief,lisa,tianqin}. In the initial phase of the mission, laser link acquisition is required for SCs of millions of kilometers away from each other before the scientific phase of the mission employing laser interferometry can take place.  
	
	In the conventional strategy\cite{Lisa_EKF,Lisa_ATP,Tianqin_ATP,Taiji_ATP}, a key payload is the CCD  camera with a wide field angle which serves to narrow the angle of the uncertainty cone from $mrad$ to $\mu rad$ and then the scan by the laser beam takes over. The problem with the CCD camera is the heat generated. Due to a lack of heat ventilation in space, a specific metallic tube structure is required to channel the heat out of SC and this complicates the payload structure. Further, the experience of the heating problem of the thrusters in LPF suggests that the thermal gradient generated will take sometime to die out\cite{QPD_ATP,QPD_ATP1,QPD_ATP2} and the scientific phase of the mission will have to wait until the thermal stability of the SC is restored. A further disadvantage in our view is that once the laser link acquisition is established, the CCD camera will become redundant and it no longer plays any  role in the scientific phase of the mission. 
	
	The aim of the present work is to look into the feasibility of replacing the CCD camera by the PAAM with an AEKF based on precision orbit determination in deep space incorporated into it. In the scientific phase, the PAAM serves to steer a laser beam to compensate for the angle generated by the relative motion between two SCs during the time it takes (around 10 seconds) for a laser beam to travel from a SC to a distant one. It also plays a role to compensate for the breathing angle at the annual level due to solar gravity\cite{PAAAEKF,breathing}. In this work, we will try to 
	understand the possible role  PAAM can play even at the stage of laser link acquisition, with the aid of an AEKF. This will avoid the heating problem generated by a CCD camera and simplifies the payload structure of SC. 
	
	The structure of this paper is organized as follows. Some background materials concerning ATP and scanning strategy are introduced in sections 2 and 3, respectively.  We begin to enter the core of our work in section 4, presenting the AEKF model with colored noise, the ATP control loop, and the noise model. Section 5 presents the simulation results and conducts an in-depth analysis of the scanning time results obtained from AEKF. In the final section, some remarks that look to the future of this work are made to conclude our work. 
	
	\newpage
	\section{Payload structure of the Laser link acquisition for inter-satellite laser interferometry.}
	\begin{figure}[hbt!]
		\centering
		\includegraphics[width=1\textwidth,height=0.4\textwidth]{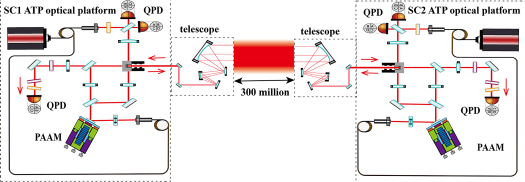}
		\caption{New schematic drawing of the laser ranging interferometry.} \label{PAAMPlace}
	\end{figure}
	Fig.~\ref{PAAMPlace} illustrates the new schematic of inter-satellite laser-ranging interferometry in our plan. In this section, we will describe the payload structure for the laser link acquisition. Apart from the PAAM that replaces a CCD camera, the two schemes share a number of payloads in common. In what follows, we shall first describe the common payloads before we go into the detailed structure of the PAAM.

	\subsection{ATP optical bench and control loop design.}

	Once a SC reaches its targeted position, precision orbit determination by the deep space network(DSN) gives an initial position with certain error margin. 
	Due to launch-induced vibrations, the lasers on the two SCs initially exhibit a frequency difference in the GHz range. To mitigate this, each SC's laser undergoes pre-stabilization, achieving a frequency stability of $30 \, \text{Hz}/\sqrt{\text{Hz}}$. 
	
	After coupling into the fiber collimator, the laser is split in a 1:99 ratio. The telescope reflect $99\%$ of the laser power. The telescope, an off-axis retroreflector composed of four mirrors\cite{telescope1}, directs the reflected beam back along the incident axis, enabling it to propagate approximately 3 million kilometers to the distant SC. The remaining $1\%$ of the laser power is directed to the fine-tracking QPD. The optical imaging system in front of the fine-tracking QPD is generally composed of either a lens system of three mirrors. Its primary function is to mitigate the impact of TTL noise, ensuring that the signal detected by the QPD reflects only the relative angular variations between the two beams. This setup effectively eliminates the influence of translational displacement between the beams.
	
	For the receiving SC, the incoming laser beam from the transmitting SC passes through the telescope and is clipped by an aperture stop at the receiving aperture. After passing through a beam splitter (BS), it is divided into two parts. One part is focused onto the acquisition QPD via an optical lens, while the other part, after passing through a dual - lens imaging system, interferes with the local SC's laser on the fine - tracking QPD. An optical phase - locked loop (OPLL) locks the phase between the transmitting and receiving SC lasers.

	\subsection{The star tracker (STR).}
	
	The STR is a key component in the attitude determination system of SC. It provides high-precision attitude information by detecting and identifying stars in its field of view (FOV). 
	Based on the in-orbit experience of the BeiDou system, the readout noise of the STR currently is at the order of $1 \times 10^{-5}$ rad. In a decade or so, it is anticipated that
	the readout noise of a STR will be improved to $1 \times 10^{-6}$ rad. By then, the PAAM together possibly with an AEKF alone is capable of fully covering the uncertain region during the scanning phase, eliminating the reliance on micro-newton thrusters for attitude adjustments\cite{thrusters}. In this work, we will focus on STR with  $1 \times 10^{-5}$ rad readout noise.

	\subsection{The role of the telescope on the acquisition phase.}
	
	The optical aperture is the primary metric for evaluating a telescope's light - gathering capability. A larger telescope aperture allows for a greater light flux, which in turn results in higher received energy. However, since the interferometric arm length is on the order of 3 million kilometers, increasing the optical aperture alone has a minimal effect on improving the received energy but would significantly increase manufacturing challenges. Drawing on the optical aperture configurations of similar telescopes in the BeiDou system, the ATP system employs an off-axis four  mirror structure telescope with an aperture of 200 mm \cite{telescope1}. During the capture phase, the telescope primarily influences the capture FOV and the scanning range of the PAAM.
	
	The telescope considered will have an FOV of 400~$\mu \mathrm{rad}$~\cite{telescope}, while the QPD, which uses DPS for angle measurement, can achieve an FOV of up to 1~$\mathrm{mrad}$. Therefore, during the capture phase, the size of the capture FOV is mainly determined by the telescope's FOV. Additionally, the telescope has a magnification of 40 times. When the PAAM is used to scan the uncertainty region, the scanning range is reduced by a factor of 40 due to the telescope's effect. Specifically, the pitch angle is reduced from $\pm 270~\mu \mathrm{rad}$ to $\pm 6.75~\mu \mathrm{rad}$, and the deflection angle is reduced from $\pm 268~\mu \mathrm{rad}$ to $\pm 6.75~\mu \mathrm{rad}$.The waist size of the emitted laser beam is comparable to the diameter of the telescope. Assuming that the beam waist radius $ r $ is approximately 20 cm, the approximate half-angle divergence of the laser can be derived using the formula for beam divergence in a Gaussian beam context:
	\begin{equation}
		\theta_{\mathrm{div}} = \frac{\lambda}{r\pi} \approx 1.69~\mu \mathrm{rad}.
	\end{equation}
	
	Due to the influence of SC pointing jitter and telescope pointing angle jitter, the half-angle of the actual effective beam receiving area can be approximated as the difference between the divergence angle and the angular jitter~\cite{telescope2}. The pointing error caused by attitude jitter and angular jitter is approximately less than $ 0.15~\mu \mathrm{rad} $, and the half-angle of the effectively received beam can be expressed as:
	\begin{equation}
		\theta_{\mathrm{effective}} = 1.54~\mu \mathrm{rad}.
	\end{equation}
	
	\subsection{PAAM-- Key distinction between the conventional strategy and the proposed one.}
	Compared to the traditional optical platform design for the acquisition phase, there are three main differences. First, a PAAM has been added to the outgoing optical path to enhance scanning efficiency during the scanning phase in coordination with micro-newton thrusters. The specific details of the scanning process will be thoroughly discussed in what follows. 
	Further, a PAAM monitoring interferometer has been incorporated into the original optical path, utilizing an optical closed-loop system to improve the pointing accuracy of the PAAM. The PAAM monitoring interferometer also enables the PAAM to suppress a portion of the pointing jitter noise introduced by the local optical platform through AEKF. This integration allows the pointing adjustments of both phases to be accomplished within a single phase. Finally, by inserting an AEKF into the PAAM, we are able to replace the CCD camera with a QPD based on DPS angle measurement technology. This substitution resolves the thermal balance issue of the optical platform and enables immediate tracking and pointing operations after the link establishment.

	The PAAM should ideally be placed at the entrance pupil of the telescope, as it alters the outgoing angle of the telescope's emitted light. Positioning the PAAM at this location constrains the size of the emitted beam, ensuring the beam's spot size within the telescope is controlled to prevent it from striking structural components and generating stray light. Additionally, an aperture is employed to further constrain the beam's spot size. By rotating the PAAM, the direction of the transmitted (TX) beam on the QPD surface, as well as the local (LO) beam direction, is adjusted.
	
	\begin{figure}[hbt!]
		\centering
		\includegraphics[width=0.4\textwidth,height=0.3\textwidth]{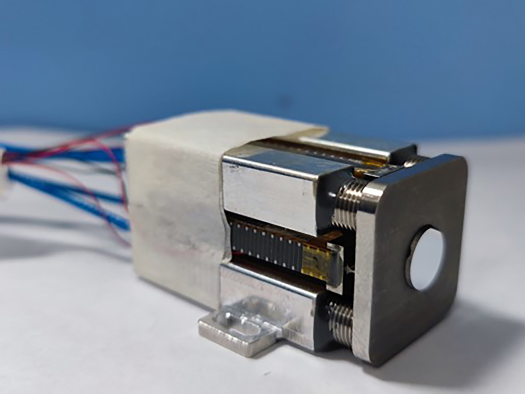}
		\caption{Physical diagram of the PAAM}
		\label{fig:image1}
	\end{figure}
	
	As shown in \mbox{Fig.\ref{fig:image1}}, PAAM as a two-dimensional motion component mounted on the optical platform that provides the PAA, is one of the key components for establishing the inter-satellite scientific interferometric link\cite{PAAM4}. It needs to address two core technical challenges: (1) Due to the extremely long arm length of the inter-satellite interferometric link, the PAA pre-pointing must achieve high-precision pointing\cite{PAAM2,PAAM3}. (2) The PAAM pointing assembly must achieve ultra-stable optical path stability, and no additional optical path noise should be introduced during its rotation. This is because the PAAM directly participates in the establishment of the scientific interferometer, and its optical path performance will directly impact the detection accuracy of the constellation \cite{PAAM,PAAM4}.The PAAM has a deflection range of $\pm270\ \mu rad$ in yaw and $\pm268\ \mu rad$ in pitch, with a pointing accuracy of $0.35\ \mu rad$ in both axes\cite{PAAM}. Considering the telescope's $30\times$ magnification factor, these ranges are reduced to $\pm9\ \mu rad$ (yaw) and $\pm8.93\ \mu rad$ (pitch), while the effective pointing accuracy is improved to $<0.012\ \mu rad$ for both axes. To guarantee operational reliability and prevent the degradation of piezoelectric actuators, during experiments, we  limit the post - telescope operational range to $\pm6\ \mu rad$ for both axes. This measure addresses two crucial constraints. First, long - term operation at maximum deflection impairs pointing accuracy. Second, extreme positions may damage the piezoelectric components. The piezoelectric actuators of the PAAM offer a rapid response ($>10\ kHz$ bandwidth in vacuum conditions). However, to alleviate thermal effects resulting from continuous high - frequency operation during the ATP phase, we set the post - telescope scanning velocity at $1\ mrad/s$.

	\subsection{PAA  calculated from orbit position and velocity information}
	At the initial stage of link establishment, the desired attitude of the SC is determined using dual-vector attitude determination based on the PAA and orbital determination errors. To calculate the desired attitude, it is necessary to convert the SC's position and velocity information in the J2000.0 coordinate system, along with the orbital determination errors, into the PAA\cite{Tang2}.
	
	\begin{figure}[hbt!]
		\centering
		\includegraphics[width=0.5\textwidth,height=0.4\textwidth]{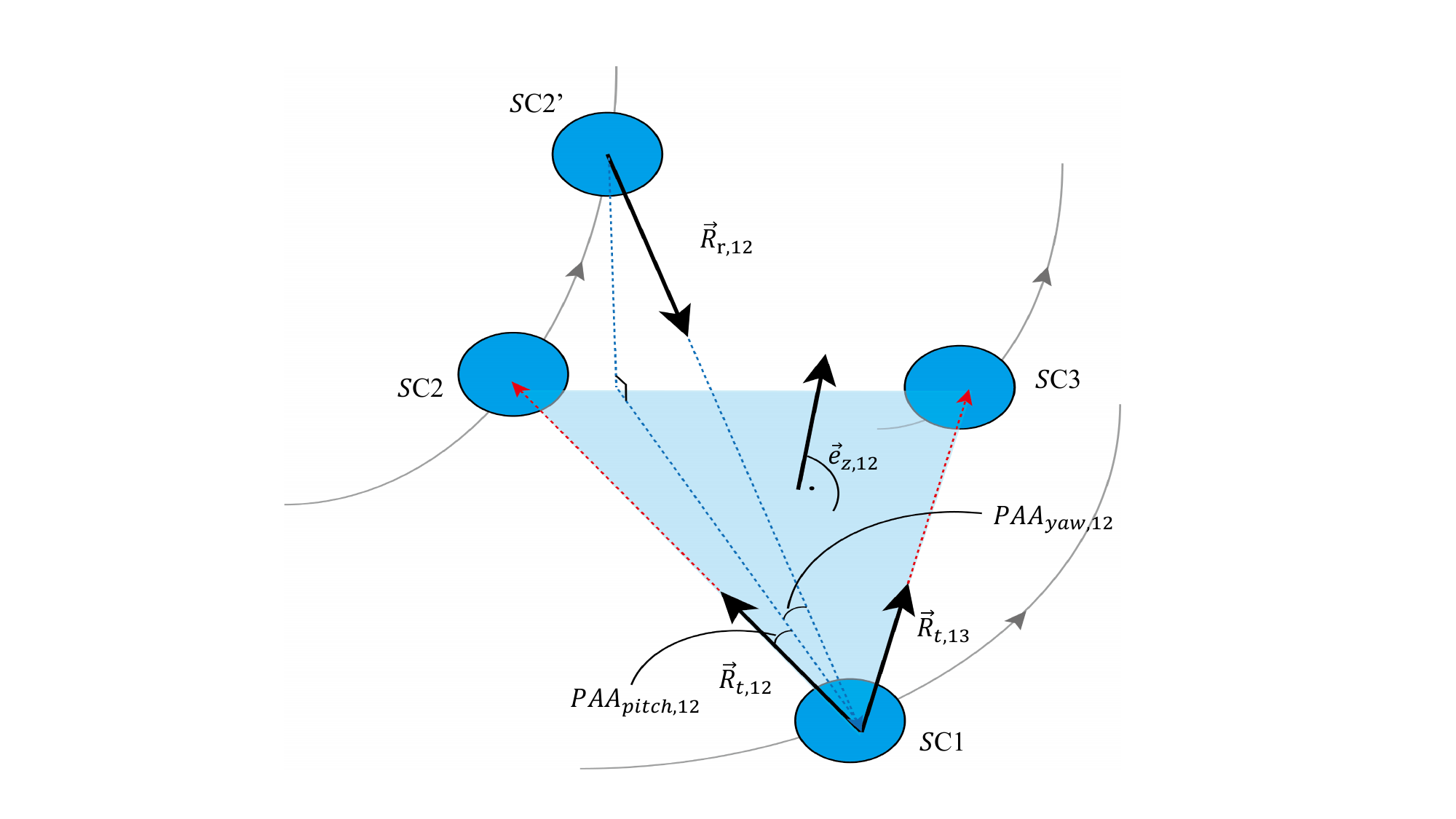}
		\caption{Definition of the Point Ahead Angle.} \label{PAASend}
	\end{figure}
	
	The PAA is defined as the angle between the direction of the transmitted beam from the local SC to the remote SC and the direction of the received beam from the remote SC to the local SC, as illustrated in Fig.~\ref{PAASend}. This angle varies annually with the orbit of the triangular constellation. Throughout this work, the J2000.0 coordinate system is adopted to describe the position and velocity of the SCs. 
	
	The positions and velocities of the $i$-th SC (where $i=1,2,3$) at a given epoch can be expressed as $[X_{i}, Y_{i}, Z_{i}]^{T}$ and $[V_{\tiny {Xi}}, V_{\tiny {Yi}}, V_{\tiny {Zi}}]^{T}$, respectively. The position and velocity errors caused by orbital determination errors are defined as $[\delta X_{i}, \delta Y_{i}, \delta Z_{i}]^{T}$ and $[\delta V_{\tiny {Xi}}, \delta V_{\tiny {Yi}}, \delta V_{\tiny {Zi}}]^{T}$. Thus, the SC position and velocity incorporating orbital determination errors can be obtained in the J2000.0 coordinate system\cite{PAAcalculate1} as.
	
	\begin{eqnarray}
		\left\{
		\begin{aligned}
			&X=[X_{i}+\delta X_{i}, Y_{i}+\delta Y_{i}, Z_{i}+\delta Z_{i}]^{T},\\
			&V=[V_{\tiny {Xi}}+\delta V_{\tiny {Xi}}, V_{\tiny {Yi}}+\delta V_{\tiny {Yi}}, V_{\tiny {Zi}}+\delta V_{\tiny {Zi}}]^{T}.
		\end{aligned}
		\right.
	\end{eqnarray}	
	The relative position and velocity of two SCs are then given by vector addition or subtraction with respect to the J2000.0 reference frame\cite{PAAAEKF}. We shall divide the PAA into two parts in the calculations: yaw and pitch, as shown in Fig.~\ref{PAASend}.
	
	$\overrightarrow{R}_{12}$ is the laser link directed from SC1 to SC2. Likewise, $\overrightarrow{V}_{12}$ is the relative velocity vector from SC1 to SC2. The beam vector $\overrightarrow{R}_{t,\ot}$ and beam vector $\overrightarrow{R}_{r,\ot}$ linking SC1 to SC2. And $\Delta t$ represents the time of laser transmission between the two SCs, and the variation caused by changes in the relative distance between the SCs is negligible. The beam vector $\overrightarrow{R}_{\small t,\scriptsize{13}}$ linking SC1 to SC3.
	
	To calculate the PAA of the SC, it is necessary to transform the origin of the coordinate system from the J2000.0 coordinate system to the center of mass of the SC. As shown in Fig.~\ref{PAASend}, we define three unit direction vectors of the SC coordinate system as 	${\overrightarrow{e}}_{x,\ot}$, ${\overrightarrow{e}}_{z,\ot}$, and ${\overrightarrow{e}}_{y,\ot }$\cite{PAAAEKF}.
	The expression of ${\overrightarrow{e}}_{z,\ot}$ is given by
	\begin{eqnarray}
		{\overrightarrow{e}}_{z,\ot}&=\frac{1}{\left|{\overrightarrow{R}}_{t,\ot}\right|^2\left|{\overrightarrow{R}}_{t,\op}\right|}
		\begin{bmatrix}
			{{e_z}_{x}}\\
			{{e}_{zy}}\\
			{{e}_{zz}}
		\end{bmatrix},
	\end{eqnarray}
	with 
	\begin{eqnarray}
		\begin{bmatrix}
			{{e_z}_{x}}\\
			{{e}_{zy}}\\
			{{e}_{zz}}
		\end{bmatrix}=
		\begin{bmatrix}
			({y_{12}+Vy_{12}\Delta t})({z_{13}+Vz_{13}\Delta t})
			-({z_{12}+Vz_{12}\Delta t})({y_{13}+Vy_{13}\Delta t}) \\
			({z_{12}+Vz_{12}\Delta t})({x_{13}+Vx_{13}\Delta t})
			-({x_{12}+Vx_{12}\Delta t})({z_{13}+Vz_{13}\Delta t}) \\
			({x_{12}+Vx_{12}\Delta t})({y_{13}+Vy_{13}\Delta t})
			-({y_{12}+Vy_{12}\Delta t})({x_{13}+Vx_{13}\Delta t})
		\end{bmatrix}.
	\end{eqnarray}
	
	The initial pointing pitch angle is then given by
	\begin{eqnarray}
		\hbox{PAA}_{\hbox{\tiny pitch,\scriptsize{12}}}
		&=&\frac{{\overrightarrow{R}}_{r,\ot}}
		{\left|{{\overrightarrow{R}}_{r,\ot}}\right|}\times {\overrightarrow{e}_{z,\ot}}\cdot \frac{{\overrightarrow{R}}_{t,12}}{\left|{\overrightarrow{R}}_{t,12}\right|}\nonumber\\
		&=& \frac{{\overrightarrow{R}}_{r,\ot}\times (\overrightarrow{R}_{t,\ot}\times\overrightarrow{R}_{ \small t,\scriptsize{13}})\cdot {\overrightarrow{R}}_{t,12}}
		{\left|{{\overrightarrow{R}}_{r,\ot}}\right|^2 \left|{\overrightarrow{R}}_{t,13}\right|\left|{{\overrightarrow{R}}_{t,\ot}}\right|}.\nonumber\\	
	\end{eqnarray}
	with
	\begin{eqnarray}
		{\overrightarrow{R}}_{r,\ot}\times (\overrightarrow{R}_{t,\ot}\times\overrightarrow{R}_{ \small t,\scriptsize{13}})\cdot {\overrightarrow{R}}_{t,12}=
		&&(({y_{12}-V_{\tiny y_{12}}\Delta t}){e}_{zz}-({z_{12}-V_{\tiny z_{12}}\Delta t}){e}_{zy})({x_{\ot }+Vx_{\ot }\Delta t})\nonumber\\
		&&+(({z_{12}-V_{\tiny z_{12}}\Delta t}){e_z}_{x}-({x_{\ot }-Vx_{\ot }\Delta t}){e}_{zz})({y_{12}+V_{\tiny y_{12}}\Delta t})\nonumber\\
		&&+(({x_{\ot }-Vx_{\ot }\Delta t}){e_z}_{y}-({y_{12}-V_{\tiny y_{12}}\Delta t}){e}_{zx})({z_{12}+V_{\tiny z_{12}}\Delta t}).
	\end{eqnarray}
	The initial pointing yaw angle is expressed as
	\begin{eqnarray}
		\hbox{PAA}_{\hbox{\tiny yaw,\scriptsize{12}}} &=&\frac{{\overrightarrow{e}}_{z,\ot } \cdot {\overrightarrow{R}}_{r,\ot}}{|\overrightarrow{R}_{r,\ot}|}\nonumber\\
		&=&\frac{1}{\left|{\overrightarrow{R}}_{t,12}\right|\left|{\overrightarrow{R}}_{t,13}\right|\left |\overrightarrow{R}_{r,\ot}\right|}({e_z}_{x}({x_{12}-V_{\tiny x_{12}}\Delta t})+{e_z}_{y}({y_{12}-V_{\tiny y_{12}}\Delta t})+{e_z}_{z}({z_{12}-V_{\tiny z_{12}}\Delta t})).
	\end{eqnarray}
	
	\subsection{The uncertainty cone and precison orbit determination in deep space. }
	\begin{figure}[hbt!]
		\centering
		\includegraphics[width=0.5\textwidth,height=0.2\textwidth]{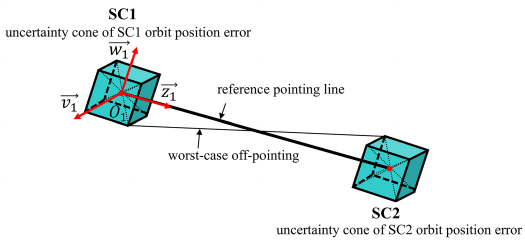}
		\caption{Uncertainty cone generated by the uncertainty of SC position in orbit determination in deep space.} \label{navigation}
	\end{figure}
	As illustrated in Fig.~\ref{navigation}, the size of the uncertainty region during the ATP phase is determined by the SC position (navigation) error introduced by the DSN. As the distance between the Earth and the triangle constellation is very similar to that between Mars and the Earth,  the experience for the precision orbit determination of the Martian mission Tianwen I will serve as a useful reference guide\cite{yangpeng}.  When the DSN performs 24-hour full-arc tracking, the orbit determination accuracy of the SC can reach 50m\cite{li2021orbit}. When the DSN tracks for 4 hours per day, the orbit determination accuracy of the SC is approximately 400m to 500m. As the DSN performs 2 hours tracking per day, the orbit determination accuracy of the SC is $2\, \text{km}$ and $0.2 \, \text{cm/s}$. 
	In an optimal scenario when the DSN performs 24-hour trajectory tracking of the SC, the link establishment can be accomplished directly using the PAAM, without the need for an AEKF.	In this work, we will consider the case when the initial orbit error is  $2\, \text{km}$ and $0.2 \, \text{cm/s}$ as input into the AEKF orbital integrator.  
	Our scheme will also work even in a less optimal worse scenario when 	
	the orbit determination accuracy of the SC is $20\, \text{km}$ and $0.2 \, \text{cm/s}$, though it takes longer scanning times for laser link establishment.
	
	Specifically, this uncertainty is characterized by the angle between the line connecting the two SCs. The following sections provide a detailed explanation of the calculation process\cite{Lisa_EKF}.
	The unit position vector from SC1 to SC2 can be expressed as
	\begin{eqnarray}
		\overrightarrow{z}_{1-2}=\frac{[x_{12},y_{12},z_{12}]^T}{L_{12,ref}},
	\end{eqnarray}
	with
	\begin{eqnarray}
		L_{12,ref}=\sqrt{x_{12}^2+y_{12}^2+z_{12}^2},
	\end{eqnarray}
	$\delta{\overrightarrow{L}_{12,err}}$ is the relative current position navigation error vector from SC1 to SC2.
	\begin{eqnarray}
		\delta{\overrightarrow{L}_{12,err}}=[\delta x_{12}, \delta y_{12}, \delta z_{12}]^T. \label{NAV_calculate1}
	\end{eqnarray}
	The projection of the relative position navigation error vector from SC1 to SC2 in the direction of the relative position vector from SC1 to SC2  can be expressed as
	\begin{eqnarray}
		\delta{\overrightarrow{L}_{12,err}}^{'}=\delta L_{12} [x_{12},y_{12},z_{12}]^T,
	\end{eqnarray}
	with
	\begin{eqnarray}
		\delta {L_{12}}=\frac{[x_{12}\delta x_{12}+y_{12}\delta y_{12}+z_{12}\delta z_{12}]}{L_{12,ref}^2}.
	\end{eqnarray}
	$\delta{\overrightarrow{r_{12}}}$ is the relative position navigation error along a coordinate axis of the inertial heliocentric frame provided by the DSN (estimated to be as 2 km).
	\begin{eqnarray}
		&&\delta{\overrightarrow{r_{12}}}=\delta{\overrightarrow{ L}_{12,err}}-\delta{\overrightarrow{ L}_{12,err}^{'}}\\ \nonumber
		&=&[\delta x_{12}-\delta L_{12} x_{12}, \delta y_{12}-\delta L_{12} y_{12}, \delta z_{12}-\delta L_{12} z_{12}]^T.
	\end{eqnarray}
	
	\begin{eqnarray}
		&&|\delta{\overrightarrow{r}_{12}}|=\sqrt{(\delta x_{12}-x_{12}\delta L_{12} )^2+(\delta y_{12}-y_{12}\delta L_{12} )^2+(\delta z_{12}- z_{12}\delta L_{12})^2}. 
	\end{eqnarray}
	The uncertainty region contribution $\theta_{u}$ from the navigation error is given by the trigonometric equation: 
	\begin{eqnarray}
		\label{worst_navigation}
		\theta_{u,12}&&=\arcsin(\frac{|\overrightarrow{\delta r_{12}}|}{L_{12,ref}-2|\overrightarrow{\delta r_{12}}|})\approx \frac{\sqrt{{\delta w}^2+{\delta v}^2}}{L_{12,ref}}\\ \nonumber 
		&&= \frac{1}{\sqrt{L_{12,ref}^2}}\sqrt{(\delta x_{12}-\delta L_{12} \cdot x_{12})^2+(\delta y_{12}-\delta L_{12} \cdot y_{12})^2+(\delta z_{12}-\delta L_{12} \cdot z_{12})^2},
	\end{eqnarray}
	where $L_{12,ref}$ is the arm length without orbit position error. \mbox{Equation \ref{worst_navigation}} represents the worst-case navigation error and it is illustrated in Fig.~\ref{navigation}.
	
	\section{Acquisition strategy and capture process.  }
	\subsection{The acquisition strategies}
	In our experimental setup for the scanning and acquisition strategy, we assume that \(\sigma_{\alpha}\) and \(\sigma_{\beta}\) are independently distributed. When two SCs reach the target positions, through orbit determination by the DSN, their coordinates are located at any point within a tube of 2 km in diameter. The SC first needs to use a star tracker for attitude adjustment to complete the initial pointing in order to avoid a back - to - back configuration. Our acquisition and scanning strategy is applicable after the SC completes the initial pointing.
	
	Successful acquisition is defined as the situation where both the FOV of the SC detector and the laser spot overlap with the target area, which ensures that the link can be established. Based on the current calculation of the detector's effective detection range, the QPD is now installed in place of the original CCD camera, and it utilizes DPS angle measurement technology. The QPD's FOV is \(\theta_{\text{FOV}} = 1.43 \, \text{mrad}\), and after being reduced by the telescope, it remains sufficient to fully cover the uncertainty region. Therefore, during the acquisition phase, the laser spot coverage on the remote SC alone serves as the criterion for successful acquisition.
	
\begin{figure}[ht!]
		\centering
		\includegraphics[width=0.4\textwidth,height=0.4\textwidth]{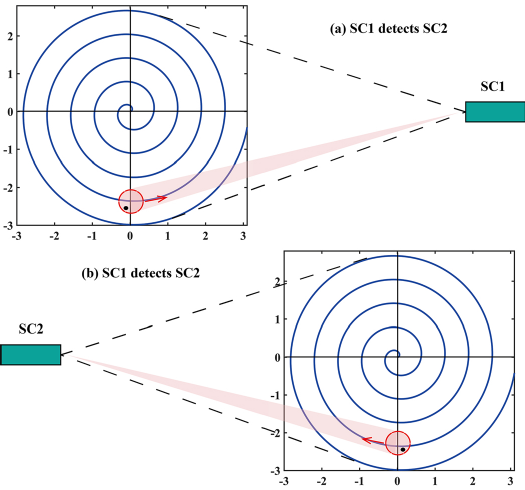}
		\caption{Schematic of the spacecraft scanning at both ends.}\label{scan}
	\end{figure}

	Based on the angular size of the uncertainty region after filtering by the AEKF, distinct acquisition strategies have been developed. To facilitate the calculation of scanning time, the AEKF-filtered results are converted into polar coordinates. Using the angular position and radius $R$ in polar coordinates, we determine the path length required for a spiral trajectory to cover a specific point as shown in  Fig.~\ref{scan}.
	The primary focus is not on the speed of link establishment but on maximizing the probability of successful link acquisition. Once a link is established, it is designed to remain stable with a low probability of disconnection. Consequently, the scanning strategy prioritizes comprehensive coverage of the uncertainty region to ensure a high success rate in capturing and establishing the link.

	\subsubsection{Strategy 1: Scan by PAAM}
	
	When the uncertain cone angle $\theta$ is greater than the laser divergence angle $\theta_{\text{div}}$ but within the dynamic range of the optical angle $\theta_{\text{\tiny{PAAM}}}$ of the PAAM, this scheme relies solely on the rapid scanning capability of the PAAM. During the scanning process, the PAAM follows an Archimedean spiral path with uniform linear velocity.
	
	\subsubsection{Strategy 2: Scanning strategy combining PAAM and micro-newton thrusters}
	
	When the uncertain cone angle $\theta$ exceeds the adjustment range of the PAAM optical angle $\theta_{\text{\tiny{PAAM}}}$, we need to expand the scanning range by combining SC attitude adjustments using micro-newton thrusters with PAAM scanning. In this scheme, both SC at the ends employ a scanning strategy that integrates discrete attitude adjustments via SC micro-newton thrusters with rapid and continuous scanning by the PAAM. The specific implementation process is as follows: The micro-newton thrusters perform discrete movements following an Archimedean spiral path with fixed step sizes and overlap ratios. For each discrete control maneuver of the micro-newton thrusters, the PAAM conducts a full-coverage scan along the Archimedean spiral path with a constant angular velocity, covering 90\% of its maximum deflection range per revolution.
	
	\subsection{Search time}
	
	In the ATP phase of the gravitational waves detection in space, a primary focus is placed on enhancing the success probability of link establishment, with an emphasis on minimizing the time required for link establishment while ensuring full coverage scanning. The experimental configuration adopts a spiral scanning method, proven to be an efficient means of encompassing the entire uncertainty region. By setting the spiral line spacing equal to the diameter of the laser beam divergence angle, full coverage of the uncertain region can be achieved during uniform linear velocity scanning. The Archimedean spiral is parameterized as 
	\begin{eqnarray}
		r=b \cdot \theta
		\label{r}.
	\end{eqnarray}
	$\theta$ represents the polar angle, which increases by $2\pi$ for every complete rotation, and $r$ denotes the radius of the helix. The parameter $b$ defines the incremental change in the helical radius per full revolution, and its relationship with the trajectory width $D_t$ is given by: $D_t = 2\pi b$. From \eqref{r}, we can calculate the total length of the spiral by integrating along the spiral path\cite{Lisa_EKF}. It is given by:
	\begin{eqnarray}
		L = \frac{1}{2} \left( r \sqrt{1 + \left( \frac{r}{b} \right)^2} + b \cdot \text{arcsinh}\left( \frac{r}{b} \right) \right).
	\end{eqnarray}
	
	When scanning the uncertainty region by means of SC attitude adjustments, the scanning process can be regarded as a uniform angular velocity scanning in which the the SC's attitude is fine-tuned through micro-newton thruster adjustments and the SC's scanning angular velocity is $\omega_{\text{micro}} = 1 \; \mu \text{rad/s}$.The scanning time $T_micro$ can be calculated using the following formula: 
	\begin{eqnarray}
		T_{micro} = \frac{L}{\omega_{\text{micro}}} 
		= \frac{1}{\omega_{\text{micro}}} \left( r \sqrt{1 + \left( \frac{r}{b} \right)^2} + b \cdot \text{arcsinh}\left( \frac{r}{b} \right) \right).
	\end{eqnarray}
	
	The scanning process can be regarded as a uniform linear velocity when only the PAAM is used to scan the uncertainty region. Currently, the scanning angular velocity of the PAAM is approximately $\omega_{\text{\tiny{PAAM}}} = 1 \; \text{mrad/s}$. The scanning time $T_1$ can be calculated using the following formula: 
	\begin{eqnarray}
		T1 = \frac{L}{\omega_{\text{\tiny{PAAM}}}} 
		= \frac{1}{\omega_{\text{\tiny{PAAM}}}} \left( r \sqrt{1 + \left( \frac{r}{b} \right)^2} + b \cdot \text{arcsinh}\left( \frac{r}{b} \right) \right).
	\end{eqnarray}
	
	In the scanning scheme in which the PAAM is coordinated with SC attitude adjustments, the scanning process can be considered as a discrete motion. Moreover, by adjusting the scanning step size and overlap ratio, full coverage of uncertain areas is maintained. Concurrently, this is paired with the uniform angular velocity scanning strategy of PAAM, allowing the scanning time, $T_{\scriptscriptstyle PAAM}$, to be calculated using the following formula:
	\begin{eqnarray}
		T_{\text{\tiny{PAAM}}} = \frac{1}{\omega_{\text{\tiny{PAAM}}}} \left( r_{\text{\tiny{PAAM}}} \sqrt{1 + \left( \frac{r_{\text{\tiny{PAAM}}}}{b} \right)^2} + b \cdot \text{arcsinh}\left( \frac{r_{\text{\tiny{PAAM}}}}{b} \right) \right),
	\end{eqnarray}
	where$r_{\text{\tiny{PAAM}}}$ is the maximum deflection angle of PAAM, $r_{\text{\tiny{PAAM}}}=6 \mu rad$.The total scan time for fourth scheme can be expressed as
	\begin{eqnarray}
		T_2 = \left\lceil\frac{r}{r_{\text{\tiny{PAAM}}}\cdot p} \right\rceil^2 \cdot T_{\text{\tiny{PAAM}}} + \left( \left\lceil\frac{r}{r_{\text{\tiny{PAAM}}}\cdot p} \right\rceil^2 - 1\right) \cdot \delta t ,
	\end{eqnarray}
	where $p$ is the overlap rate in the discrete scanning process of SC attitude adjustment by micro-newton thrusters, $p=0.8$. $\delta t$ is the time needed to change the direction between adjacent spots. 
	
	\subsection{The capture process}
	During the capture phase, the light received from the distant SC is weak, making it challenging for the local detector to observe the distant light spot when the local laser is active. Therefore, at the beginning of the scanning process, a toggling approach is employed: when the local laser is on, the distant receiving SC must turn off its laser. Conversely, when the distant SC activates its laser for scanning, the local laser must be switched off. The inter-satellite link is established once the QPD for scientific measurements on each SC detects interference signals.
	
	In this process, when using PAAM for scanning, it is necessary to return the PAAM to its original position by adjusting the SC's attitude to avoid the constraints of PAAM's adjustment range during the scientific measurement phase. Specifically, within the detector's field of view, as the PAAM moves towards the zero position, the micro-newton thrusters adjust the SC's attitude in the opposite direction. Throughout this adjustment process, it is crucial to synchronize the PAAM adjustment speed with that of the micro-newton thrusters to prevent disruptions to the interference signal on the QPD.

	\section{Adaptive Extended Kalman Filter for ATP}
	In this section, we shall introduce the AEKF method, on the basis of which we develop a new control algorithm for the ATP phase. We will briefly describe the AKEF framework\cite{PAAAEKF} and then  apply it to the laser link acquisition in what follows. 
	
	\subsection{Construction of an adaptive extended Kalman Filter}%
	Consider a hybrid extended Kalman filter in which the physical system concerned is governed by continuous and nonlinear dynamic equations, and the measurements are discrete in time. 
	\begin{eqnarray}
		\left\{\begin{array}{l}\overset{.}{X} = f\left( {X,t} \right) + w(t),\label{dynamic1}
			\\Z_{k} = h_{k}\left(X_{k},v_{k} \right)\label{measurement},
		\end{array}\right.
	\end{eqnarray}
	where both the dynamic function $f\left( {X,t} \right)$ and the measurement function $h_{k}\left( X_{k} \right) $ are nonlinear, $w(t)$ is the continuous noise. 
	$w_{k}$ is the system noise and $v_{k}$ is the colored measurement noise.
	\begin{eqnarray}
		\left\{\begin{array}{l} E[{w(t)w^{T}\left( {t + \tau} \right)}]= Q_{c}\delta(t),
			\\w_{k}\sim(0,Q_{k}),
			\\Q_{k} = Q_{c}(k\Delta t)/\Delta t,
			\\v_{k + 1} = {\Psi_{k ,k-1}}v_{k} + \xi_{k}\label{dynamic4}.	
		\end{array}\right.
	\end{eqnarray}
	
	We use Gaussian white noise for system noise in this simulation and the covariance matrix of $w_{k}$ is $Q_{c}$. At time $k$, an average $Q_{k}$ is taken for the continuous $Q_{c}$ at time $t$, and the obtained average $Q_{k}$ represents the covariance matrix of the process noise at time $k$. In the formula, $\xi_{k}$ is the white noise sequence with the mean of 0, and $\Psi_{k,k-1}$ is the coefficient transfer matrix of the colored measurement noise.
	
	In the AEKF, we can treat the measurement noise as a state quantity and include it in the state equation\cite{wangyan1}. The state equation of the Kalman filter, which includes the extended system noise and the measurement noise, may be written as\mbox{equation \ref{statement}}.
	\begin{eqnarray}
		\begin{bmatrix}
			X_{k + 1} \\
			v_{k + 1} \\
		\end{bmatrix} = \begin{bmatrix}
			\phi_{k,k - 1} & 0\\
			0 & \Psi_{k ,k-1} \\
		\end{bmatrix} \begin{bmatrix}
			X_{k} \\
			v_{k} \\
		\end{bmatrix} + \begin{bmatrix}
			\Gamma_{k + 1,k}  & 0 \\
			0 & I\\
		\end{bmatrix}\begin{bmatrix}
			W_{k}  \\
			\xi_{k} \\
		\end{bmatrix}\label{statement},
	\end{eqnarray}
	where $\Gamma_{k + 1,k}$ is an identity matrix. The formula of EKF algorithm subject to the colored measurement noise can be referred to\cite{PAAAEKF}.
	
	\subsection{ATP control loop design}
	\begin{figure}[hbt!]
		\centering
		\includegraphics[width=0.7\textwidth,height=0.3\textwidth]{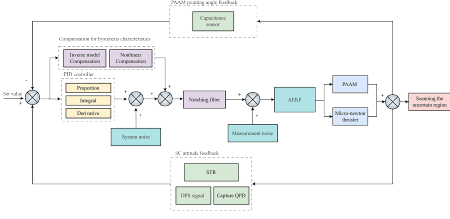}
		\caption{ATP control framework.} \label{controlloop}
	\end{figure}	
	The AEKF method is an efficient predictive filtering approach that utilizes data from the previous step to predict the outcome of the next step. In a LISA-type mission, unlike the EKF considered before for clock synchronization purposes in the pre-TDI data post-processing \cite{wangyan2,wangyan3}, the AEKF for PAAM will be carried out on orbit. Our design of the ATP control mainly includes the beam pointing control of PAAM and the control of the SC's attitude by the micro-newton thruster. In the PAAM beam pointing control system, the FPGA control chip receives the feedback signals from the capacitance sensors of PAAM. In the micro-newton thruster-based SC attitude control system, the thrusters regulate the SC's attitude by using the feedback signals from STRs and the far-field spot position data of the target SC, which the QPD provides based on DPS angle measurement. 
	
	The ATP control system combines the information of the SC orbit integrator for feedforward control. Fig \ref{controlloop} shows the block diagram design of the control system of the ATP. In this control system, a theoretical model is established by AEKF and SC orbital integrator to control the PAAM. The output value controlled by the PID is weighted with the system noise as the state input to AEKF, and the measurement noise is added to simulate the actual situation on the SC. The whole control loop is closed-loop controlled by capacitance sensors, STRs, and a QPD based on DPS angle measurement technology. Taking into account the creep and hysteresis of piezoelectric ceramics in the PAAM, nonlinear compensation, and notching filters are added to the control loop to improve the stability of the whole control loop.
	
	\subsection{AEKF model for ATP}
	First, we define a 20-dimensional column state vector, which includes the position and velocity information of three SCs\cite{wangyan1,PAAAEKF}.
	\begin{eqnarray}
		x =[\overrightarrow{x_{1}},\overrightarrow{x_{2}},\overrightarrow{x_{3}},\overrightarrow{v_{1}},\overrightarrow{v_{2}},\overrightarrow{v_{3}},\overrightarrow{\delta x_{1}}, \overrightarrow{\delta x_{2}}, \overrightarrow{\delta x_{3}},\overrightarrow{\delta v_{1}}, \overrightarrow{\delta v_{2}}, \overrightarrow{\delta v_{3}}]^T,
	\end{eqnarray}
	where $\overrightarrow{x_{i}}=(x_{i},y_{i},z_{i})^T$ are the SC positions, $\overrightarrow{v_{i}}=(v_{xi},v_{yi},v_{zi})^T$ are the SC velocities, $\overrightarrow{\delta {x_{i}}}=(\delta x_{i},\delta y_{i},\delta z_{i})^T$ and $\overrightarrow{\delta {v_{i}}}=(\delta v_{xi},\delta v_{yi},\delta v_{zi})^T$ are the position error and velocity error in SC orbit determination, and $i = 1; 2; 3$ is the SC index.
	
	The dynamics of a single SC is described by the Keplerian equation for planetary motion given by 
	\begin{eqnarray}
		\ddot{\overrightarrow{x_{k}}} = \sum_{p}GM_{p}\frac{\overrightarrow{x_{k}}-\bm{R}_{p}}{\mid \overrightarrow{x_{k}}-\bm{R}_{p}\mid^{3}},
	\end{eqnarray}
	where $\overrightarrow{X_{k}}$ is the position of a SC, $M_{p}$, $\bm{R}_{p}$ are the mass and the coordinates of the pth celestial body (the Sun and the planets) in the solar system, ${\overrightarrow{X_{k}}-\bm{R}_{p}}$ is a vector pointing from that SC to the $p$ th celestial body. Gravitational forces originated from  the Sun and the major planets including Mercury, Venus, Earth+Moon, Mars, Jupiter, Saturn, Uranus, and Neptune \cite{Tang2} are considered. 
	
	The dynamic equation can be written as
	\begin{eqnarray}
		\frac{d}{dt}
		\begin{bmatrix}
			\overrightarrow{x_{k}}\\
			\overrightarrow{v_{k}}\\
		\end{bmatrix}
		&=f\left(\overrightarrow{x_{k}},\overrightarrow{v_{k}}\right)
		&=	\begin{bmatrix}
			\overrightarrow{v_{k}}\\
			\sum_{p}GM_{p}\,\frac{\overrightarrow{x_{k}}-\bm{R}_{p}}{\mid \overrightarrow{x_{k}}-\bm{R}_{p}\mid^{3}}\\
		\end{bmatrix}.
	\end{eqnarray}
	Define $\alpha =\left(\overrightarrow{x_{k}},\overrightarrow{v_{k}}\right)^T$\cite{wangyan1}, then we have
	\begin{eqnarray}
		\phi=\frac{\partial f}{\partial \alpha}=
		\begin{bmatrix}
			O_3 & I_3 \\
			A   & O_3 
		\end{bmatrix}.
	\end{eqnarray}
	Here $O_3$ is the zero matrix, $I_3$ is the $3\times3$ identity matrix\cite{wangyan2}, and the expression of the $3\times3$ matrix A is given as:
	\begin{eqnarray}
		A	&=&-\sum_{p}\frac{GM_{p}}{\mid \overrightarrow{x_{k}}-\bm{R}_{p}\mid^{3}}\mathbf{I}_3
		+ \sum_{p}\frac{3GM_{p}}{\mid \overrightarrow{x_{k}}-\bm{R}_{p}\mid^{5}}\,(\overrightarrow{x_{k}}-\bm{R}_{p})(\overrightarrow{x_{k}}-\bm{R}_{p})^T.
	\end{eqnarray}
	
	\begin{eqnarray}
		\theta_{u,12}
		=\frac{\sqrt{(\delta x_{12})^2+(\delta y_{12})^2+(\delta z_{12})^2}}{\sqrt{x_{12}^2+y_{12}^2+z_{12}^2}}.
	\end{eqnarray}
	
	\begin{eqnarray}
		\frac{d}{dt}
		\begin{bmatrix}
			\overrightarrow{\delta x_{k}}\\
			\overrightarrow{\delta v_{k}}
		\end{bmatrix}
		&=f\left(\overrightarrow{\delta x_{k}},\overrightarrow{\delta v_{k}}\right)
		&=	\begin{bmatrix}
			\overrightarrow{\delta v_{k}}\\
			\sum_{p}GM_{p}\,\frac{\overrightarrow{\delta x_{k}}-\bm{R}_{p}}{\mid \overrightarrow{\delta x_{k}}-\bm{R}_{p}\mid^{3}}
		\end{bmatrix}.
	\end{eqnarray}
	For the entire system, the dynamic matrix $\phi=\frac{\partial f}{\partial x}$ is $30\,\times\,30$. We omit its explicit expression here, as it can be obtained in a straightforward way from the above formulae.
	In our work that follows, we simulate the PAA data of 10 days. Since the PAA changes very slowly with one year periodicity, we set the sampling frequency of the  AEKF to 1s. In the initial design process, we have considered the influence of SC displacement, velocity, and acceleration on PAA, but in our subsequent simulation, we find that the relative acceleration of the SC is irrelevant to the PAA calculations. 
	
	Next, we shall present the two-dimensional measurement equation. The measurement equation which links up the positions and velocities of the three SCs and the yaw and pitch PAAs. In   our scheme, the position error in the navigation may be written as 
	\begin{eqnarray}
		z_{k} &= h_{k}(x_{k}, v_{k}) 
		= \begin{bmatrix}
			\text{PAA}_{\text{yaw, 12}}, & \text{PAA}_{\text{pitch, 12}}, & \theta_{\text{u,12}}, \\
			\text{PAA}_{\text{yaw, 21}}, & \text{PAA}_{\text{pitch, 21}}, & \theta_{\text{u,21}}
		\end{bmatrix}
	\end{eqnarray}
	where $\hbox{PAA}_{\hbox{\tiny pitch,\scriptsize{ij}}}$ and $ \hbox{PAA}_{\hbox{\tiny yaw,\scriptsize{ij}}} $ are respectively the pitch and the yaw PAA between SCi and SCj.  ${\theta}_{\hbox{\scriptsize{U,ij}}}$ is the uncertainty cone caused by worst-case navigation error between SCi and SCj. $v_{k}$ is the measurement nosie. $H_{k}$ is a $6×36$ dimensional observation matrix. We omit its explicit expression here, since it can be obtained straightforwardly from the formulae below. The element $H_{k}[i,j]$ in the matrix $H_{k}$ may be expressed as:
	\begin{eqnarray}
		H_{k}[i,j]=\frac{\partial z_{k}[i]}{\partial x_{k}[j]}.
	\end{eqnarray}
	As an example, the [1,1] component of $H_{k}$, with the step index $k$ omitted, may be given as follows.	
	\begin{eqnarray}
		\begin{bmatrix}
			{{e_z}_{x}}\\
			{{e}_{zy}}\\
			{{e}_{zz}}
		\end{bmatrix}=
		\begin{bmatrix}
			({y_{12}+Vy_{12}\Delta t})({z_{13}+Vz_{13}\Delta t})
			-({z_{12}+Vz_{12}\Delta t})({y_{13}+Vy_{13}\Delta t}) \\
			({z_{12}+Vz_{12}\Delta t})({x_{13}+Vx_{13}\Delta t})
			-({x_{12}+Vx_{12}\Delta t})({z_{13}+Vz_{13}\Delta t}) \\
			({x_{12}+Vx_{12}\Delta t})({y_{13}+Vy_{13}\Delta t})
			-({y_{12}+Vy_{12}\Delta t})({x_{13}+Vx_{13}\Delta t})
		\end{bmatrix}.
	\end{eqnarray}
	\begin{eqnarray}
		H[1,1]
		&=&\frac{1}{\left|{\overrightarrow{R}}_{t,12}\right|^2\left|{\overrightarrow{R}}_{t,13}\right|^2\left |\overrightarrow{R}_{r,\ot}\right|^2}\nonumber\\
		&&[({e_z}_{x}+(({z_{12}+Vz_{12}\Delta t})
		-({z_{13}+Vz_{13}\Delta t}))({y_{12}-V_{\tiny y_{12}}\Delta t})+(({y_{13}+Vy_{13}\Delta t})
		-({y_{12}+Vy_{12}\Delta t}))({z_{12}-V_{\tiny z_{12}}\Delta t}))\nonumber\\ %
		&&(\left|{\overrightarrow{R}}_{t,12}\right|\left|{\overrightarrow{R}}_{t,13}\right|\left |\overrightarrow{R}_{r,\ot}\right|)-({e_z}_{x}({x_{12}-V_{\tiny x_{12}}\Delta t})+{e_z}_{y}({y_{12}-V_{\tiny y_{12}}\Delta t})+{e_z}_{z}({z_{12}-V_{\tiny z_{12}}\Delta t}))\nonumber\\
		&&(\left|{\overrightarrow{R}}_{t,13}\right|\left |\overrightarrow{R}_{r,\ot}\right|\frac{\partial \left|{\overrightarrow{R}}_{t,12}\right|}{\partial X_1}+\left|{\overrightarrow{R}}_{t,12}\right|\left |\overrightarrow{R}_{r,\ot}\right|\frac{\partial \left|{\overrightarrow{R}}_{t,13}\right|}{\partial X_1}+\left|{\overrightarrow{R}}_{t,12}\right|\left|{\overrightarrow{R}}_{t,13}\right|\frac{\partial \left |\overrightarrow{R}_{r,\ot}\right|}{\partial X_1})]
	\end{eqnarray}
	with
	\begin{eqnarray}
		\frac{\partial \left|{\overrightarrow{R}}_{t,12}\right|}{\partial X_1}
		= \frac{2({x_{12}+V_{\tiny x_{12}}\Delta t})}{2\left|{\overrightarrow{R}}_{t,12}\right|}
		= \frac{({x_{12}+V_{\tiny x_{12}}\Delta t})}{\left|{\overrightarrow{R}}_{t,12}\right|}
	\end{eqnarray}
	\begin{eqnarray}
		\frac{\partial \left|{\overrightarrow{R}}_{r,12}\right|}{\partial X_1}
		= \frac{2({x_{12}-V_{\tiny x_{12}}\Delta t})}{2\left|{\overrightarrow{R}}_{r,12}\right|}=
		\frac{({x_{12}-V_{\tiny x_{12}}\Delta t})}{\left|{\overrightarrow{R}}_{r,12}\right|}
	\end{eqnarray}
	\begin{eqnarray}
		\frac{\partial \left|{\overrightarrow{R}}_{t,13}\right|}{\partial X_1}= \frac{2({x_{13}+V_{\tiny x_{13}}\Delta t})}{2\left|{\overrightarrow{R}}_{t,13}\right|}= \frac{({x_{13}+V_{\tiny x_{13}}\Delta t})}{\left|{\overrightarrow{R}}_{t,13}\right|} 
	\end{eqnarray}
	
	In standard practice, the coefficient transfer matrix $\Psi_{k,k-1}$ of colored measurement noise is determined by the ARMA model. In the present context, the definition of $\Psi_{k,k-1}$ is relatively simple and can be considered as a special case of the ARMA model in which the autoregressive parameter is 1, and the moving average parameter is 0 \cite{ARMR}. We have also tried to use more sophisticated  ARMA model to estimate the value of the $\Psi_{k,k-1}$ matrix, but the results are not as good as the methods used here. As shown in state equation of AEKF, we only consider the colored noise component in the measurement noise and did not include white noise. Therefore, the $\xi_{k}$ term can be directly ignored. As shown in \mbox{\ref{Psi}}, we approximately take the ratio of the measurement noise amplitudes at the previous moment and the current moment as input to the coefficient transfer matrix $\Psi_{k,k-1}$ of the colored measurement noise. The simple choice here is likely due to a very slow variation of the colored measurement noise in the time domain at the annual level. This colored measurement noise is a linear superposition of all the colored noise types that we will discuss in detail shortly. 
	\begin{eqnarray}
		{\Psi_{k,k - 1}} =v_{k}/ v_{k - 1}\label{Psi}.	
	\end{eqnarray}
	
	In a standard AEKF, the size of the $Q_{k}$ and $R_{k}$ matrix is automatically adjusted by observing the prediction error and its mean square error matrix and introducing the fading factor to obtain a good estimation state \cite{adaptiveEKF}. The AEKF designed in this paper is slightly different from the traditional AEKF. In the AEKF designed by us, the covariance matrix of measurement noise $R_k$ is updated in real-time according to the magnitude of measurement noise, while the covariance matrix of system noise $Q_k$ is updated every month or so, based on the accuracy of orbit prediction.
	
	\begin{figure}[hbt!]
		\centering
		\includegraphics[width=0.8\textwidth,height=0.45\textwidth]{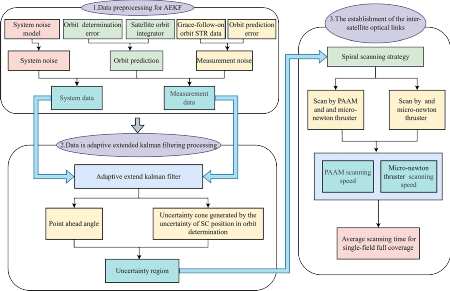}
		\caption{ATP flow chart based on AEKF.} \label{ATP calculation}
	\end{figure}
	In this study, the flowchart for ATP based on AEKF is illustrated in Fig.~\ref{ATP calculation}. During the first step of data preprocessing, we rely on orbit prediction to acquire the SC's position and velocity information.
	The system noise data is linearly superimposed onto the orbit prediction data to form the system data. The orbit prediction data, linearly superimposed with the Grace-follow-on orbit STR noise data, constitutes the measurement data input into the AEKF.    In the second step, during the AEKF process, the orbit prediction information needs to be updated weekly. Through the filtering results of AEKF, we can calculate the size of the initial uncertain region. Finally, by considering the scanning speeds of PAAM and micro-newton thrusters, we calculated the average scanning time for single-field full coverage under the two scanning strategies and conducted statistical analysis on the experimental results at the same time. 
	
	\subsection{Noise model}
	
	\subsubsection{System Noise Model}
	To develop a system model for the AEKF, we established a system noise model for the ATP phase.  This noise model primarily includes solar pressure noise, noise generated by SC attitude jitter and PAAM dynamic actuation error noise. We totally added a position error of 2km and a velocity error of 0.2cm/s to the SC as the system noise. We incorporate the SC's position and velocity errors as the initial state inputs of the orbital integrator in the AEKF, thus establishing the system model for the AEKF. Additionally, the covariance matrix $Q_{k}$ is configured in accordance with the intensity of the system noise.
	
	\subsubsection{Measurement noise model}
	
	This section primarily elaborates on the ATP noise model, serving as the foundation for quantifying measurement noise in our subsequent simulations. During the simulation, we linearly superimpose the position and velocity data of the SC provided by the orbital integrator with the orbit determination error data generated by random noise. This approach is used to simulate the orbital prediction information. We consider this information as the true value of AEKF and linearly superimpose it with the measurement noise data, which serves as the input of measurement value for AEKF. At the same time, the covariance matrix $R_k$, that characterizes the measurement noise within the AEKF, is dynamically computed from the generated noise and updated in real-time as shown in \mbox{\ref{R_update}}\cite{adaptiveEKF}.
	\begin{eqnarray} 
		R_{k} = cov(v_{k})=E\{v_{k}{v_{k}}^T\}\label{R_update}.
	\end{eqnarray}
	
	The measurement noise model can be divided into two main components: the measurement noise model for the PAA and the measurement noise model for the  $\theta_{\mathrm{u}}$. For the PAA measurement noise model, we primarily derive it using amplitude transformations based on on-orbit data from the GRACE Follow-On mission. For the  $\theta_{\mathrm{u}}$ measurement noise model, white noise is used to simulate orbital prediction errors. This chapter will provide a detailed introduction to each of these models.
	
	\paragraph{PAA measurement noise model}\
	
	The measurement noise model for PAA accounts for various sources and  mainly includes piston noise of PAAM, SC attitude jitter noise, laser intensity noise, laser shot noise, STR read-out noise, detector equivalent input current noise and other optical platform noises. During the ATP phase, the readout noise of the STR is the primary source of measurement noise. In our simulation, a STR with  \(1 \times 10^{-5}\) rad readout noise was selected. To ensure that our simulation closely mirrors the actual conditions in orbit, we applied amplitude transformation to the STR measurement data from Grace-Follow-on mission, using it as the input for the measurement noise model. 
	
	First, we need to convert the STR quaternion data into the Euler angles representing the telescope's vector direction. The attitude of the SC body relative to the ground inertial coordinate system is defined by the attitude quaternion. The attitude quaternion $q_{ib} \in R^{4 \times 1}$is represented as $q_{ib} = \left\lbrack {q_{ib,0},q_{ib,1},q_{ib,2},q_{ib,3}} \right\rbrack^{T}$. 
	To convert the four-element attitude data of the STR into the direction of the telescope vector, the four-element data must be converted into the rotation matrix $R_T$\cite{quaternions}.
	\begin{figure}[hbt!]
		\centering
		\includegraphics[width=0.4\textwidth,height=0.3\textwidth]{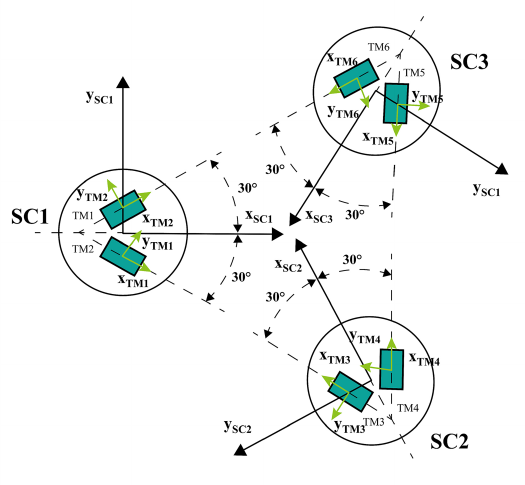}
		\caption{Spacecraft coordinates and telescope coordinates.} \label{coordinates}
	\end{figure}	
	As shown in Fig.~\ref{coordinates}, there is a 30 degrees angle between the STR position vector and the telescope exit direction vector, so the STR position vector must be rotated 30 degrees around the Z axis. If the STR vector direction is expressed as $[1  0  0]^T$, then the telescope exit direction vector can be expressed as
	\begin{eqnarray}
		\begin{bmatrix}
			\alpha_{ij} \\
			\beta_{ij} \\
			\gamma_{ij}
		\end{bmatrix} =\begin{bmatrix}
			1 \\ 0 \\ 0
		\end{bmatrix}*\begin{bmatrix}
			\cos\left( \frac{\pi}{6} \right) & {\sin\left( \frac{\pi}{6} \right)} & 0 \\
			{- {\sin\left( \frac{\pi}{6} \right)}} & {\cos\left( \frac{\pi}{6} \right)} & 0 	\\
			0 & 0 & 1
		\end{bmatrix}*R_T=
		\begin{bmatrix}
			{\left( {\sqrt{3}\left( {{q_{ib,0}}^{2} + {q_{ib,1}}^{2}} \right) - \sqrt{3}/2} \right) + \left( {- q_{ib,0}{*q}_{ib,3} + q_{ib,1}{*q}_{ib,2}} \right)} \\
			{\sqrt{3}\left( {q_{ib,0}{*q}_{ib,3} + q_{ib,1}{*q}_{ib,2}} \right) + \left( {\left( {{q_{ib,0}}^{2} + {q_{ib,2}}^{2}} \right) - 1/2} \right)} \\
			{\sqrt{3}\left( {{- q}_{ib,0}{*q}_{ib,2} + q_{ib,1}{*q}_{ib,3}} \right) + \left( {q_{ib,0}{*q}_{ib,1} + q_{ib,2}{*q}_{ib,3}} \right)}
		\end{bmatrix}.
	\end{eqnarray}
	
	The error introduced during the conversion of STR quaternion data to the Euler angles representing the telescope's vector direction ranges from $10^{-19} \, \text{rad}$ to $10^{-9} \, \text{rad}$\cite{quaternions}. This level of accuracy is sufficient to meet our requirements. From this, we can derive the conversion relationship between the telescope vector and the STR quaternion data.
	
	\begin{figure}[hbt]
		\centering
		\includegraphics[width=0.45\textwidth,height=0.3\textwidth]{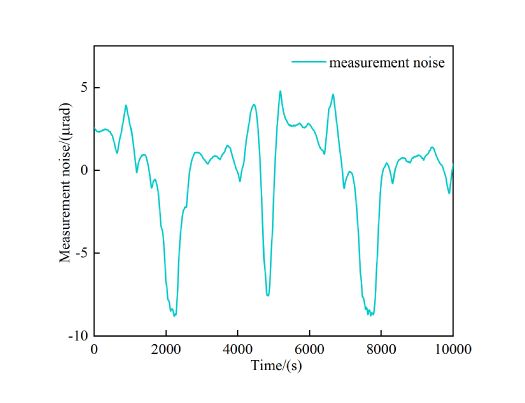}
		\caption{Measuremnet noise.} \label{PAA_measurement_noise}
	\end{figure}
	
	Subsequently, by applying amplitude variations to the time-domain noise data, we obtain time-domain PAA measurement noise data with average amplitudes of $1 \times 10^{-5}\; \hbox{rad}$, as shown in Fig.~\ref{PAA_measurement_noise}.
	\paragraph{$\theta_{\mathrm{u}}$ measurement noise model}\
	To work out a $\theta_{\mathrm{u}}$ measurement model for AEKF, consider
	\begin{eqnarray}
		\left\{
		\begin{aligned}
			&\delta X=[\delta X_{i},\delta Y_{i},\delta Z_{i}]^{T},\\
			&\delta V=[\delta V_{\tiny {Xi}},\delta V_{\tiny {Yi}},\delta V_{\tiny {Zi}}]^{T},
		\end{aligned}
		\right.
	\end{eqnarray}		
	where ${\delta X}$ and $\delta V$ are the errors in the position and velocity, respectively, in the orbit prediction. We take the SC's position and velocity data generated by the orbit integrator as input to the AEKF while incorporating the orbit prediction error in the form of measurement noise. Accordingly, \ref{NAV_calculate1} to \ref{worst_navigation} are then employed to calculate the uncertainty region caused by navigation errors.
	
	Based on the orbital prediction experience from Tianwen-I, when a 24-hour orbit determination is conducted using the DSN, the velocity determination error is $0.2 \, \text{cm/s}$, and the position determination error is $200 \, \text{m}$. Under these conditions, the establishment of the inter-satellite optical link can be achieved without relying on AEKF. However, from the viewpoint of autonomous navigation, it is feasible to use orbit prediction data for a month in the measurement model of the AEKF, provided the required precision in orbit determination on the position and velocity of the SC in orbit are respectively within the margins of 50km and 2cm/s.This uncertainty region is input as the source of measurement noise for $\theta_{\mathrm{u}}$, resulting in an error margin of approximately $10 \, \mu \text{rad}$. We can also update the orbit determination error by ground tracking during data transmission between the SC and the ground station. Our proposed scheme is primarily designed for scenarios with lower orbit determination accuracy. The measurement noise of $\theta_{\mathrm{u}}$ is added to the actual $\theta_{\mathrm{u}}$ value provided by the orbital integrator in the form of white noise.
	
	\section{Results and discussion}
	In this section, we conduct an error analysis on the results of PAA and  $\theta_{\mathrm{u}}$ after AEKF filtering. Additionally, based on the filtered results, we linearly superimpose the installation error and measurement noise of the STR. This allows us to investigate the single-field scan time changes under different scanning strategies, both before and after AEKF filtering.
	
	\subsection{AEKF result analysis.}
	\begin{figure}[hbt!]
		\centering
		\includegraphics[width=0.55\textwidth,height=0.4\textwidth]{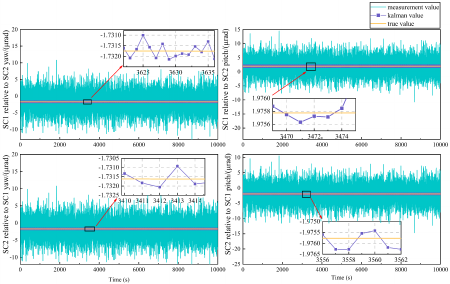}
		\caption{The AEKF prediction results of PAA between SC1 and SC2.} \label{sc12AEKF}
	\end{figure}
	We simulated measurements of about 10,000 seconds with a sampling frequency of 1 Hz.In this simulation, only the laser link between two of the three SCs was considered for analysis.
	Fig.~\ref{sc12AEKF} shows the AEKF prediction results of the PAA between SC1 and SC2. Prior to AEKF filtering, the average PAA error between SC1 and SC2 was 9~$\mu$rad. After AEKF filtering, the maximum AEKF prediction error of SC1 relative to SC2 in the pitch direction is $1.1\hbox{nrad}$, and the maximum AEKF prediction error of  SC1 relative to SC2 in the yaw direction is $ 1.2\hbox{nrad}$. The maximum AEKF prediction error of SC2 relative to SC1 in the pitch direction is $ 1.01 \hbox{nrad}$, and the maximum AEKF prediction error of SC2 relative to SC1 in the yaw direction is $ 1.21\hbox{nrad}$.
	
	\begin{figure}[hbt!]
		\centering
		\includegraphics[width=0.5\textwidth,height=0.4\textwidth]{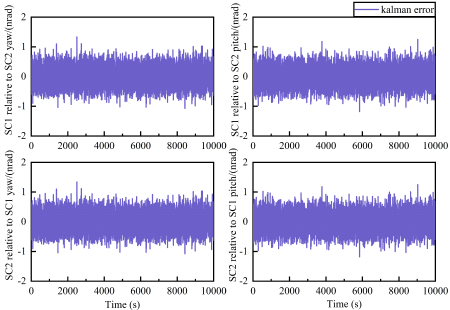}
		\caption{The AEKF prediction errors of PAA between SC1 and SC2.} \label{sc21AEKF}
	\end{figure}
	Fig.~\ref{sc21AEKF} shows the AEKF prediction errors of the PAA between SC1 and SC2. The average prediction error of the AEKF from SC1 relative to SC2 in the pitch direction is $ 0.69 \hbox{nrad}$, and the average prediction error of the AEKF from SC1 relative to SC2 in the yaw direction is $ 0.67\hbox{nrad}$.  The average prediction error of the AEKF from SC2 relative to SC1 in the pitch direction is $ 0.67\hbox{nrad}$, and the average prediction error from SC2 relative to SC1  in the yaw direction is $ 0.76 \hbox{nrad}$. The AEKF filtering effectively decreased the average PAA error of SC1 relative to SC2 by 82.31~dB in the pitch direction and 82.56~dB in the yaw direction, while the average navigation error of SC2 relative to SC1 was reduced by 82.56~dB in the pitch direction and 81.47~dB in the yaw direction.
	\newpage
	\begin{figure}[hbt!]
		\centering
		\includegraphics[width=0.8\textwidth,height=0.36\textwidth]{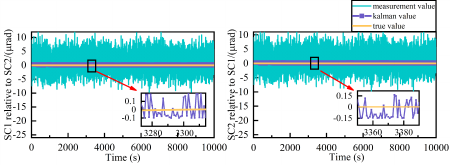}
		\caption{ $\theta_{\mathrm{u}}$ AEKF prediction results.} \label{UR-WCNE}
	\end{figure}
	Fig.~\ref{UR-WCNE} presents the prediction results of the navigation error between SC1 and SC2 using the AEKF during the scanning process. Prior to AEKF filtering, the average navigation error of SC1 relative to SC2 was 9~$\mu$rad. After AEKF filtering, the maximum navigation error of SC1 relative to SC2 decreased to 0.27~$\mu$rad, with an average navigation error of 0.1~$\mu$rad. Similarly, before AEKF filtering, the average navigation error of SC2 relative to SC1 was 10~$\mu$rad, which was reduced to a maximum of 0.23~$\mu$rad and an average of 0.16~$\mu$rad after filtering. The AEKF filtering effectively decreased the average navigation error of SC1 relative to SC2 by 39.08~dB, while the average navigation error of SC2 relative to SC1 was reduced by 35~dB.
	\newpage
	\begin{table}[h]
		\centering
		\caption{AEKF errors between SC1 and SC2.}
		\begin{tabular}{cccc}
			\toprule
			&  & \textbf{SC1 relative to SC2 } & \textbf{SC2 relative to SC1 } \\
			\midrule
			{\textbf{Max noise}} & Yaw & 1.06088E-05 & 1.02847E-05 \\
			& Pitch & 1.17175E-05 & 1.13103E-05 \\
			& $\theta_{u}$ & 1.13103E-05 & 1.12847E-05 \\
			\midrule
			{\textbf{Average noise}} & Yaw & 9.06088E-06 & 9.03847E-06 \\
			& Pitch & 9.07175E-06 & 9.01103E-06 \\
			& $\theta_{u}$ & 9.03103E-06 & 1.02847E-05 \\
			\midrule
			{\textbf{AEKF max error}} & Yaw & 1.1E-09 & 1.01E-09 \\
			& Pitch & 1.2E-09 & 1.21E-09 \\
			& $\theta_{u}$ & 0.27E-06 & 0.23E-06 \\
			\midrule
			{\textbf{AEKF average error}} & Yaw & 0.67E-09 & 0.76E-09 \\
			& Pitch & 0.69E-09 & 0.67E-09 \\
			& $\theta_{u}$ & 0.1E-06 & 0.16E-06 \\
			\bottomrule
		\end{tabular}
		\label{table1}
	\end{table}
	\mbox{Table \ref{table1}} provides a summary of the maximum and average total noise levels under colored measurement noise conditions, as well as the maximum and average prediction errors for PAA and  $\theta_{\mathrm{u}}$ using the AEKF. The results presented in Table~\ref{table1} indicate that the AEKF significantly suppresses noise under colored measurement noise conditions. Additionally, noise suppression is more effective in the pitch direction compared to the yaw direction. This difference is likely due to the PAA having a smaller adjustment range in the pitch direction compared to the yaw direction, leading to a smoother variation process that enables the AEKF to achieve more effective noise suppression.
	
	\begin{table}[h]
		\centering
		\caption{ Fitting result.}
		\begin{tabular}{ccccc}
			\toprule
			& \textbf{Yaw (SC1 to SC2)} & \textbf{Pitch (SC1 to SC2)} & \textbf{Yaw (SC2 to SC1)} & \textbf{Pitch (SC2 to SC1)}  \\ 
			\midrule
			\textbf{SSE} & 8.965E-16 & 8.983E-16 & 8.764E-16 & 8.433E-16\\ 
			\textbf{RMSE} & 1.751E-08 & 1.953E-08 & 1.684E-08 & 1.734E-08\\ 
			\textbf{R-square} & 0.9689 & 0.9688 & 0.9690 & 0.9691  \\ 
			\textbf{Adjusted R-square} & 0.9688 & 0.9684 & 0.9687 & 0.9688  \\ 
			\bottomrule
		\end{tabular}
		\label{table3}
	\end{table}
	In our simulation, the orbital dynamics and updates to the SC's state of motion are incorporated into the iterations of the AEKF. \mbox{Table \ref{table3}} summarizes the results of the maximum prediction errors for the pitch PAA and yaw PAA between the two SCs, both before and after filtering. 
	\newpage
	\begin{figure}[hbt!]
		\centering
		\includegraphics[width=0.7\textwidth,height=0.5\textwidth]{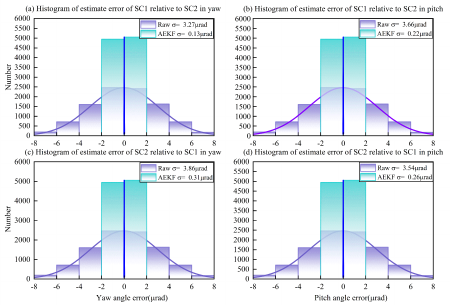}
		\caption{Histogram of the pitch and yaw error distribution between SC1 and SC2 before and after AEKF filtering.} \label{histogram}
	\end{figure}
	Fig.~\ref{histogram} shows the histogram of the pitch and yaw error distribution for SC1 relative to SC2 and SC2 relative to SC1, both before and after AEKF filtering. The AEKF effectively reduces the pitch and yaw errors of SC1 relative to SC2 and SC2 relative to SC1, thereby decreasing the size of the initial uncertainty region during the ATP phase. Furthermore, the AEKF has a more pronounced effect on reducing the RMS of the pitch and yaw direction errors.
	\newpage
	\subsection{Comparison of single-field full coverage scanning results}
	\begin{figure}[hbt!]
		\centering
		\includegraphics[width=0.7\textwidth,height=0.35\textwidth]{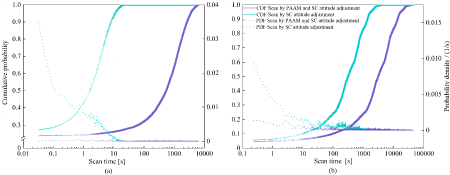}
		\caption{(a)Cumulative probability curves and probability density curves against the scanning time after the application of AEKF.(b)Cumulative probability curves and probability density curves against the scanning time before the application of AEKF.} \label{scantime1}
	\end{figure}
	
	Fig.~\ref{scantime1} show the cumulative probability curves and probability density curves before and after the application of AEKF under different scanning strategies. The results demonstrate that the scanning method integrating PAAM scanning with SC attitude adjustment can save more scanning time than the method that solely relies on SC attitude adjustment. Meanwhile, under different scanning strategies, AEKF filtering is effective in reducing the scanning time.
	
	\begin{table}[htbp]
		\caption{\label{table4} Average scanning time comparison before and after AEKF filtering.}
		\centering
		\begin{tabular}{ c c c }
			\hline
			~ & scan by PAAM and micro-newton thruster  & scan by micro-newton thruster  \\
			\hline
			Raw & 448.5717s & 1601.7s  \\
			\hline
			After AEKF & 8.6153s &  541.6171s  \\
			\hline
		\end{tabular}
	\end{table}	
	Table \ref{table4} summarizes the average scanning time, both before and after AEKF filtering, required for single-field full coverage. The comparison includes two types of scenarios. One scenario involves scanning using a combination of the PAAM and micro-newton thrusters, while the other relies solely on micro-newton thrusters. The results indicate that AEKF filtering effectively reduces the time needed for full coverage scanning. Furthermore, employing the PAAM in conjunction with micro-newton thrusters significantly enhances the efficiency of the link-establishment process. As a result, it greatly reduces the overall link acquisition time compared to the traditional strategy that relies solely on micro-newton thrusters.
	
	\section{Concluding remarks}
	
	Our  work presents a new laser link acquisition scheme in the detection of gravitational waves in space. In place of the CCD camera in the conventional scheme,  an AEKF is incorporated into the control loop of the PAAM to steer a laser beam in such a way to reduce the uncertainty cone in the initial acquisition. Further, the heating together  with the heat ventilation problem of the CCD camera in the conventional scheme is avoided. At the same time, the new scheme seamlessly integrates the coarse and fine acquisition processes in the conventional scheme into a single step. This enhances in a significant way the acquisition efficiency while reducing operational time. Numerical simulations in scenarios closely resemble the prospective on orbit situations  further verify the feasibility of this scheme. 
	
	Currently, we are setting up  a tabletop experiment to validate the AEKF constructed in the hardware setup. In addition, we are also looking at the feasibility to enlarge the dynamic range of the PAAM so as to  replace micro-newton thruster scanning functions entirely during the ATP phase. More experimental results will be reported soon. It is also expected that the new scheme will  provide new alternatives for establishing laser link in  future inter-satellite laser communication systems in deep space.
	
	\begin{acknowledgments}
		This work is supported by the National Key R$\&$D Program of China (2022YFC2203701).
	\end{acknowledgments}

	
	\clearpage
	\newpage
	\vspace{155mm}
	
	\nocite{*}
	\bibliography{aps}
	
\end{document}